\documentclass[12pt]{article}
\usepackage{epsfig}
\usepackage[hang,bf,small]{caption}
\DeclareGraphicsExtensions{.eps.gz,.eps,.ps,.ps.gz}
\parskip=\itemsep               

\newcommand{\beq}{\begin{equation}}
\newcommand{\eeq}{\end{equation}}
\newcommand{\beqar}[1]{\begin{eqnarray}\label{#1}}
\newcommand{\eeqar}{\end{eqnarray}}

\newcommand{\as}{\alpha_S}

\def\arnps#1#2#3{  {\it Ann. Rev. Nucl. Part. Sci. }{\bf #1} (19#2) #3}
\def\npb#1#2#3{    {\it Nucl. Phys. }{\bf B#1} (19#2) #3}

\def\prd#1#2#3{    {\it Phys. Rev. }{\bf D#1} (19#2) #3}

\def\zpc#1#2#3{    {\it Z. Phys. }{\bf C#1} (19#2) #3}

\relax
\begin{document}
\title{
{\Large \bf   Scaling Phenomena from Non-Linear Evolution }\\\
{ \Large \bf   in  High Energy DIS}}
\author{
{ M. ~Lublinsky\thanks{e-mail: mal@techunix.technion.ac.il}~~$\mathbf{{}^{a)}}$}\\[4.5ex]
{\it ${}^{a)}$  Department of Physics}\\
{\it  Technion -- Israel Institute of   Technology}\\
{\it  Haifa 32000, ISRAEL}\\[4.5ex]
}

\maketitle
\thispagestyle{empty}
                      
\begin{abstract} 
The numerical solutions of the non-linear evolution equation are shown to display the
 ``geometric'' scaling recently discovered in the experimental data. The phenomena hold
both for proton and nucleus targets
for all $x$ below $10^{-2}$ and $0.25\, {\rm GeV^{2}}\le Q^2 \le 2.5\times10^3\,  {\rm GeV^{2}}$.
The scaling is practically exact (few percent error) in the saturation region. In addition,  
 an  approximate scaling is  found   in the validity domain of the linear evolution where
it holds with about 10\% accuracy.

Basing on the scaling phenomena we determine the saturation scale $Q_s(x)$ and study
both its $x$-dependence and the atomic number dependence for the nuclei.
 \end{abstract}
\thispagestyle{empty}
\begin{flushright}
\vspace{-16.5cm}
\today
\end{flushright}   
\newpage
\setcounter{page}{1}

\section{Introduction}

The experimental data on the
structure function $F_2$  were recently discovered to display exciting  
phenomena called the ''geometric'' scaling \cite{scalingexp}. 
Namely,  the total $\gamma^*p$ cross section is not a function
of two independent variables $x$ and $Q$ but rather a function of a single variable 
$\tau=Q/Q_s(x)$. The  function $Q_s(x)$ is a new scale called the saturation scale.
The scaling  holds experimentally with 10\%
accuracy in the whole kinematic region of $x\le 10^{-2}$ \cite{scalingexp}.
 
These remarkable phenomena require theoretical explanation. In fact,
the scaling behavior is actually anticipated in high density QCD and is strongly related to the
appearance of the saturation scale \cite{BL,MV,LT,LT1}. 
During the interaction a   parton cascade is developed.
 When the parton density (we mean the packing factor) is not large
the transverse momenta of the partons   are strongly ordered. Such a system  evolves 
according to the linear  DGLAP equation \cite{DGLAP} which describes the gluon emission.
As a result of this radiation  the number of partons rapidly increases. 
However in the high  parton density phase
annihilation processes become significant and they  suppress the gluon radiation resulting
in the saturation of the density. This scenario happens at the  saturation scale $Q_s(x)$
\cite{MV,GLR,MUQI},  which has a meaning of an average transverse momentum of partons 
in the cascade. At  photon virtualities below $Q_s(x)$ the  ordering in the momenta does not 
persist any more and all partons have the same momenta $Q_s(x)$.  
This is a domain where the  evolution cannot be described by a linear equation.
 A  non-linear evolution should be used instead. 
The scaling is a property of this kinematic
region. It just says that the very same  average momentum $Q_s(x)$ in the cascade 
can be approached from two directions either by varying the virtuality $Q$ at fixed $x$ or
vise versa. The scaling phenomena are expected at $Q < Q_s(x)$.
Furthermore, the saturation scale  $Q_s(x)$ can be defined as a scale at which
the scaling  breaks down.

As a result of the above  discussion we conclude that the scaling phenomena should
be a consequence of the non-linear evolution  whith the non-linear effects switching on at 
the saturation scale.  
Numerous theoretical efforts  to   understand theoretically 
 the mechanisms responsible for the parton saturation 
 \cite{MV,GLR,MUQI,SAT,ELTHEORY,BA,KO,ILM} lead 
finally to  the very same nonlinear evolution equation  
\cite{GLR,MUQI,BA,KO,ILM,MU94,Braun}. This equation
 has a most transparent form in the color dipole approach:

\setcounter{equation}{0}
\begin{eqnarray}
\label{EQ} 
   N({\mathbf{x_{01}}},Y;b)&\,=\,&N({\mathbf{x_{01}}},Y_0;b)\, {\rm exp}\left[-\frac{2
\,C_F\,\as}{\pi} \,\ln\left( \frac{{\mathbf{x^2_{01}}}}{\rho^2}\right)(Y-Y_0)\right ]\,
+\nonumber  \\ & & \frac{C_F\,\as}{\pi^2}\,\int_{Y_0}^Y dy \,  {\rm exp}\left[-\frac{2
\,C_F\,\as}{\pi} \,\ln\left( \frac{{\mathbf{x^2_{01}}}}{\rho^2}\right)(Y-y)\right ]\,\times
\\ \int_{\rho} \,& d^2 {\mathbf{x_{2}}} & 
\frac{{\mathbf{x^2_{01}}}}{{\mathbf{x^2_{02}}}\,
{\mathbf{x^2_{12}}}} \nonumber 
\left(\,2\, N({\mathbf{x_{02}}},y;{ \mathbf{ b-
\frac{1}{2}
x_{12}}})- N({\mathbf{x_{02}}},y;{ \mathbf{ b -
\frac{1}{2}
x_{12}}}) N({\mathbf{x_{12}}},y;{ \mathbf{ b- \frac{1}{2}
x_{02}}})\right) \nonumber
\end{eqnarray}

The equation is written for $N(r_{\perp},x; b) = Im
\,a^{el}_{dipole}(r_{\perp},x; b)$ whith $a^{el}_{dipole}$ being the elastic  amplitude
for   a dipole of size $r_{\perp}$ scattered at the impact parameter $b$. 
The  rapidity $Y=-\ln x$ and $Y_0=-\ln x_0$. The
ultraviolet cutoff $\rho$
is needed to regularize the integral, but it does not appear in physical quantities. In the large
$N_c$ limit (number of colors)   $C_F=N_c/2$.

The equation (\ref{EQ}) describes  the following physical picture. The evolution kernel
$\frac{\mathbf{x^2_{01}}}{\mathbf{x^2_{02}}\,\mathbf{x^2_{12}}}$ is a probability for 
the dipole of size $\mathbf{x_{10}}$ to decay 
into two dipoles of  sizes $\mathbf{x_{12}}$ and $\mathbf{x_{02}}$. Then these two dipoles 
 interact independently with the target (linear  term in the equation). The nonlinear 
term in the evolution   takes into account
the Glauber corrections for the interaction. These corrections are due to the screening
between the two dipoles, and hence they contribute with the negative sign.

 It can be seen 
from the form of the equation  (\ref{EQ}) that it contains no information about target. 
The only dependence on a target  is coded in the initial conditions of the evolution  
at some initial value $x_0$. 
We take for the initial conditions the Glauber-Mueller (GM) formula,
 which is proven to be correct initial
distribution   for  nuclear  targets \cite{KO}. 
For the proton target we also use the Glauber - Mueller
formula. However, in this case the procedure is less justified theoretically and we rely in our 
choice on the fact that this formula describes well the experimental data at not too low $x$. 
The initial conditions are:
\beq
\label{ini}
 N(\mathbf{x_{01}},x_0;b)\,=\,N_{GM}(\mathbf{x_{01}},x_0;b)\,,
\eeq
with 
\beq
\label{Glauber}
N_{GM}(\mathbf{x_{01}},x;b)\,=\,1\,\,-\,{\rm exp}\left[ - \frac{\as \pi  \mathbf{x^2_{01}}}
{2\,N_c\, R^2}\,xG^{DGLAP}(x,  4/\mathbf{x_{01}^2})\,S(\mathbf{b})\right]\,.
\eeq
The equation (\ref{Glauber}) takes into 
 account  the  multiple  dipole-target interaction in the eikonal 
approximation \cite{DOF3,DOF1,DOF2}. The gluon density $xG^{DGLAP}$ is a solution
of the DGLAP equation \cite{DGLAP}.
The function $S(b)$ is a  dipole  profile function inside the target, while $R$ stands for 
its radius. 

Solutions to the equation (\ref{EQ}) were studied in asymptotic limits in Refs. \cite{LT,LT1}. 
Numerical solutions of  the  equation  (\ref{EQ}) 
were reported in the Refs. \cite{Braun,LGLM,Braun2,LL}. 
We continue studying the properties  of the solutions obtained
in the Refs. \cite{LGLM, LL}.  In the present  work we concentrate on the scaling phenomena
displayed by the solutions of (\ref{EQ}). 
 Indeed, it was shown in the Ref. \cite{LT1} that in the double logarithmic
approximation, the solutions of the equation (\ref{EQ})  scale with a good accuracy in a wide 
high energy region. The recent paper \cite{Braun2} reports on the numerical observation
of the scaling phenomena.

The main goal of the present paper is to discover  the scaling phenomena 
in the solutions  for both  proton and nuclei obtained  in the Refs. \cite{LGLM,LL}. Basing
on this property we determine the saturation scale $Q_s(x)$ and for nuclei we study its 
$A$ dependence. 

The paper is organized as follows. The next section (2)
 is devoted to the scaling phenomena and saturation scale for the proton. 
The scaling on nuclei is investigated in section 3. The final section (4) concludes the work.

\section{Scaling  and saturation scale for proton} 

In the recent paper \cite{LGLM} the nonlinear evolution equation (\ref{EQ}) was solved 
numerically for  constant value of the strong coupling constant $\as=0.25$.  
The goals of the present research are in further study of the physical properties
displayed by the solutions of (\ref{EQ}).

Compared to the Ref.  \cite{LGLM} in the present paper we slightly modified the 
large distance behavior of the initial conditions at $x=10^{-2}$. In fact, no information about
large distances is known. The GRV parameterization \cite{GRV}
entering the initial conditions ends up
at distances $\simeq 0.5$ fm.  In the previous paper  \cite{LGLM} the extrapolation 
to larger distances was done by a constant, which does  not approach unity at the 
very large distances. Moreover, such initial conditions cannot be consistent with a scaling 
being a purely dynamical property of the  evolution equation. 
There exists a transition region below
$x=x_0=10^{-2}$ where the solutions of (\ref{EQ}) are sensitive to the initial conditions. In this
transition region we do not expect to observe any scaling phenomena. The transition region
is estimated to end up at $x\simeq 10^{-4}$. Below $x\simeq 10^{-4}$ 
the initial conditions are forgotten
and the dynamics is governed  by the pure evolution. Then the scaling sets in and indeed it
 is seen in the solutions obtained  in the Ref.  \cite{LGLM}.  
 
In order to eliminate the transition region, in the present
work we extrapolate the large distance behavior of the initial conditions consistently with
the asymptotics and the scaling.  The following procedure can be suggested. We take a solution
of  Ref. \cite{LGLM} at some $x$ well bellow the transition region, say at $x=10^{-6}$.
Appropriately rescaled this solution is used for the large distance extrapolation of the initial 
conditions at $x=10^{-2}$. The above described  improvement of the initial conditions 
modifies slightly the large distance behavior of the 
solution in the transition region   $10^{-4}\le x\le 10^{-2}$ and restores
the scaling in this region.
Note that in a sense our procedure already implies the scaling at very large 
distances.

\begin{figure}[htbp]
 \epsfig{file=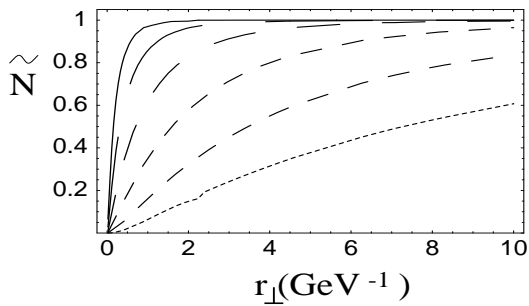,width=70mm, height=40mm}
\begin{minipage}{9.5 cm}
\vspace{-4.5cm}
 \caption[]{\it Solutions of the equation (\ref{EQ}) as a function of 
distance.  The different curves correspond to solutions at $x=10^{-7}$ 
(the upper curve), $10^{-6}$ and so on  down
to $x=10^{-2}$ (the lowest curve)}
\label{sol}
\end{minipage}
\end{figure}

The solutions obtained $\tilde N(r_\perp,x)\equiv N(r_\perp,x;b=0)$
 are displayed on the Fig. (\ref{sol}). The different
curves correspond to the different values of $x$.
As can be seen from the
Fig. (\ref{sol}), at any fixed $x$ 
the solution $\tilde N$ behaves in a step like manner as a function of distance: 
at small distances it tends to zero, 
while at larger distances the saturation value unity is approached. 

\subsection{Scaling phenomena}
In this section we study a possible scaling behavior of the solution $\tilde N$. 
As has been mentioned, the double logarithmic approximation of the solutions of 
the master equation (\ref{EQ})  \cite{LT1} as well as  general analyses of similar non-linear 
equations \cite{BL} predict this new scaling
phenomenon in the saturation region $r_{\perp} > 1/Q_s(x)$. In the saturation region  
this scaling implies  the dipole-target amplitude to be
 a function of only one variable $\tau = r^2_{\perp} \cdot Q^2_s(x)$:
\beq 
\label{SCALING}
\tilde N(r_\perp,x)\, \equiv \,\tilde{N}(\tau)\,.
\eeq
Indeed, with proper rescaling of the variable
$r_\perp$ all the curves in the Fig. (\ref{sol}) can be mapped one onto another. 
This is a manifestation of the scaling property (\ref{SCALING}).
  
A rigorous numerical procedure for the scaling  detection can be defined.
It is  useful  to introduce  the rapidity variable $y=\ln 1/x$. 
 Let us define the following derivative functions assuming the 
scaling behavior (\ref{SCALING}):
\beq 
 Ny(r_\perp,x)\,\equiv\,-\,\frac{\partial \tilde N}{\partial y}\,=\,
  \frac{d \tilde N}{d \tau}\,r^2_\perp\,\frac{d \,Q_s^2(x)}{d \ln x}\,.
\label{DX}
\eeq
\beq
 Nr(r_\perp,x)\,\equiv\, r_\perp^2\,\frac{\partial \tilde N}{\partial r_\perp^2}\, =\,
 \frac{d \tilde N}{d \tau}\,r^2_{\perp}\, Q^2_s(x) \,.
\label{DR}
\eeq

From the equations (\ref{DX}) and (\ref{DR}) one can see that the ratio $N_y/N_r$
is a function of  only one variable $x$:
\beq
 Ra(r_\perp,x)\,\equiv\, \frac{ Ny}{ Nr}\,=\,\frac{d \ln Q_s^2(x)}{d \ln x}\,.
\label{Ra}
\eeq

\begin{figure}[htbp]
\begin{tabular}{c c}
 \epsfig{file=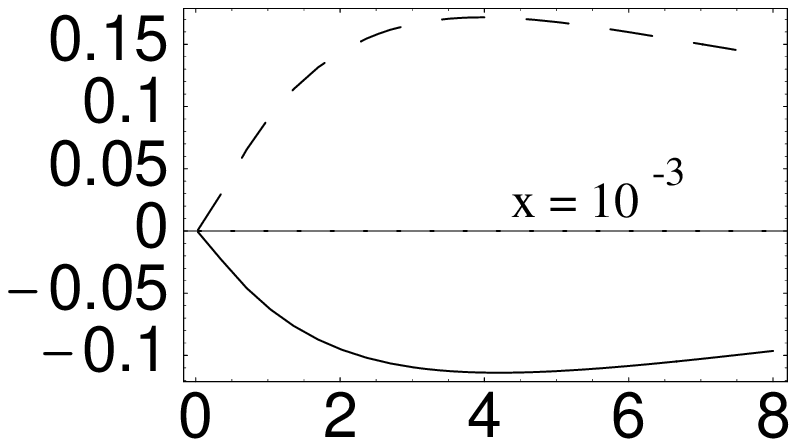,width=70mm, height=40mm}&
\epsfig{file=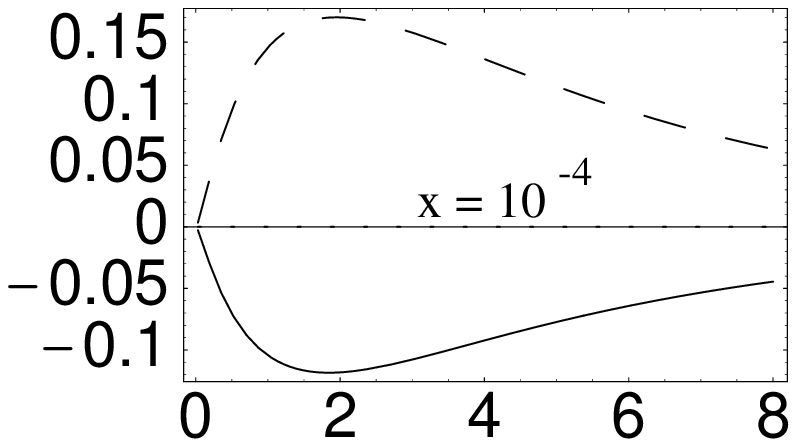,width=70mm, height=40mm}\\ 
 \epsfig{file=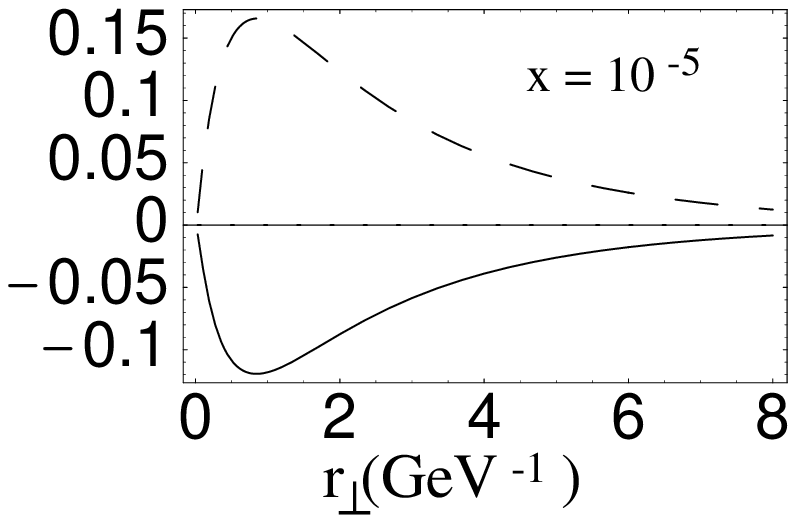,width=70mm, height=42mm}&
\epsfig{file=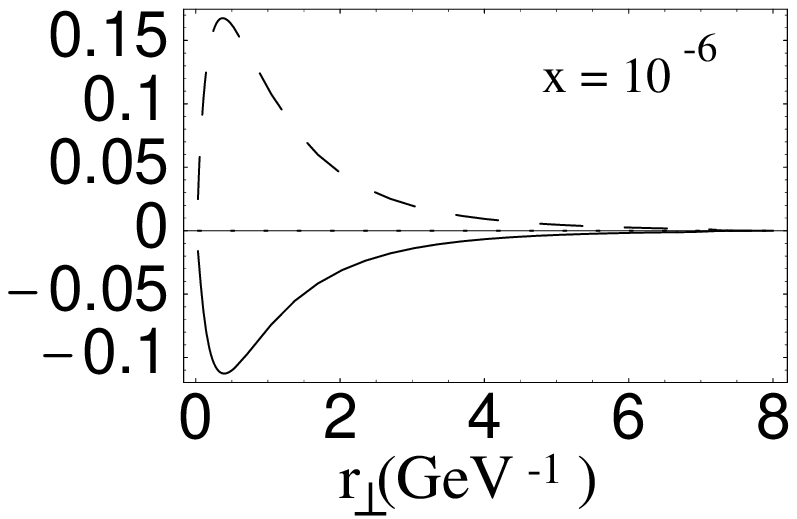,width=70mm, height=42mm}\\ 
\end{tabular}
  \caption[]{\it The derivative functions $Nr$ (dashed line) and $Ny$ (solid line) 
as functions of the distance $r_\perp$. }
\label{deriv}
\end{figure}

Our goal is to investigate the above ratio from the solutions obtained.
In the Fig. (\ref{deriv}) the derivative functions  $Nr$ and $Ny$ are plotted versus $r_\perp$ 
for various values of $x$. One can clearly observe similarity in the behavior of  these
 functions. This is actually a  sign of the scaling phenomena. 
Both functions $Nr$ and $Ny$ possess extremum points  at which the  derivatives
with respect  to $r_\perp$  vanish.  If the scaling behavior  takes place then it follows
from (\ref{SCALING}) that both  $Nr$ and $Ny$ are extreme at the same points. In fact,
 this condition is clearly  observed with a very good accuracy (Fig. \ref{deriv}). 

In order to establish the scaling phenomena numerically  
we have to check if the function $Ra$ 
is indeed  $r_\perp$ independent. However, it is clear that we cannot expect exact 
numerical independence. So, a numerical criteria  for the scaling existence has to be defined. 
In this paper we study scaling within the distance interval 
$0.04\, {\rm GeV^{-1}}\le r_\perp \le 10 \, {\rm GeV^{-1}}$ that corresponds to  
$0.25\, {\rm GeV^{2}}\le Q^2 \le 2.5\times 10^3\,  {\rm GeV^{2}}$.
Since the experimental accuracy for the scaling is about 10\% we define the following
condition for its acceptance:
\beq
\label{cond}
\frac{{\rm max}\,  \Delta Ra}{ {\rm max}\, \Delta N_{r,y}}\, \le\, 10\%,
\eeq
where  ${\rm max}\,  \Delta Ra$ is a maximal variation (in percents) within the interval of interest 
of the function $Ra$ with the distance $r_\perp$ at fixed $x$. The functions 
${\rm max}\, \Delta N_{r,y}$ are similarly defined. The condition (\ref{cond}) says that 
we accept for the scaling some  small $r_\perp$ dependence  of the function $Ra$ in a scale
of large variations of the functions $Nr$ and $Ny$. 

\begin{figure}[htbp]
\begin{tabular}{c c}
 \epsfig{file=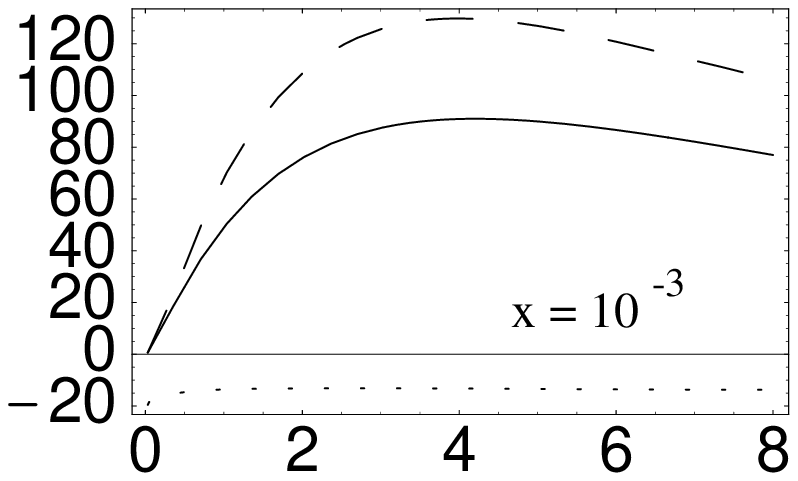,width=70mm, height=38mm}&
\epsfig{file=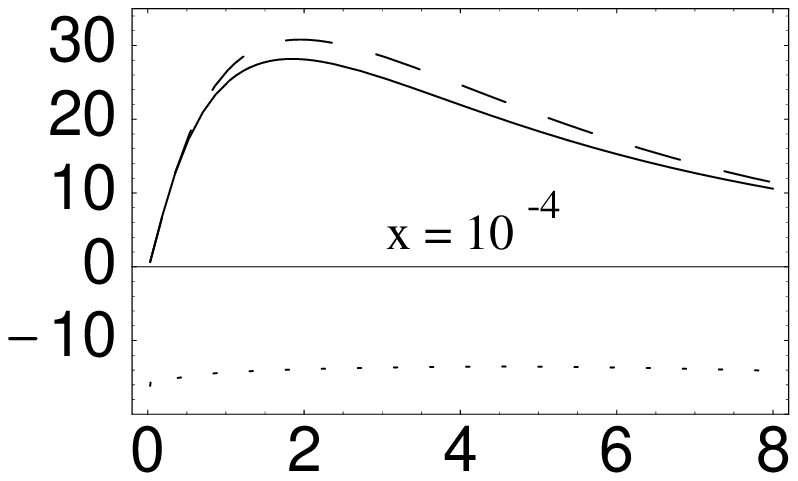,width=70mm, height=38mm}\\ 
 \epsfig{file=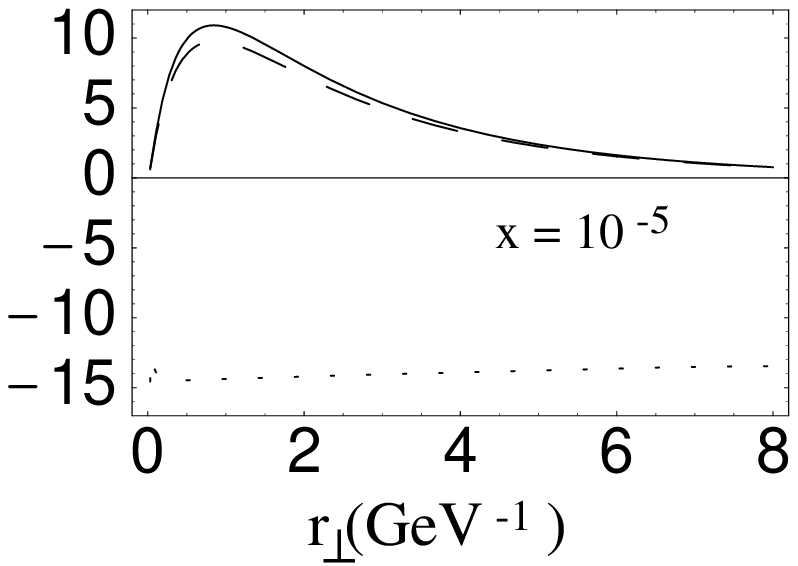,width=70mm, height=42mm}&
\epsfig{file=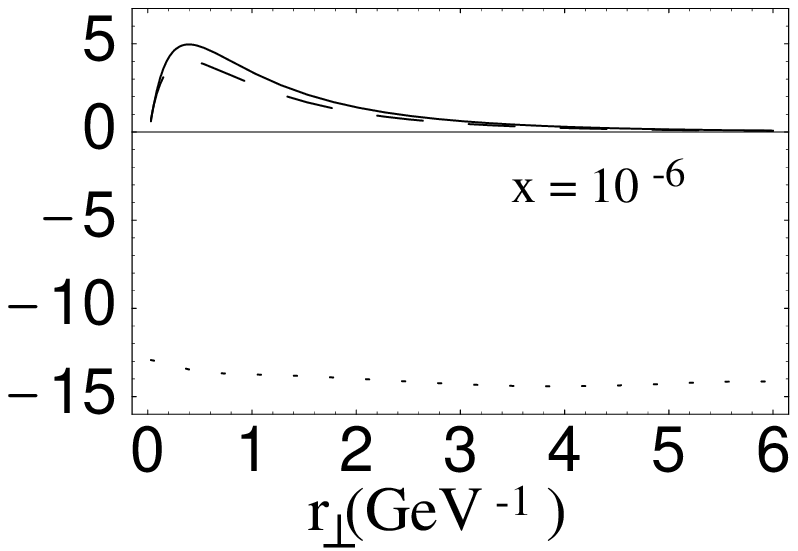,width=70mm, height=42mm}\\ 
\end{tabular}
  \caption[]{\it The  scaling   as a function of the distance  $r_\perp$. 
The positive curves
  are $Nr/Nr_{min}$ (dashed line) and   $Ny/Ny_{min}$ (solid line). The dotted line is 
$20\times Ra$.}
\label{ratio}
\end{figure}

The figure (\ref{ratio}) presents the main results. The three lines correspond to 
functions $Nr$ and $Ny$ divided by their minimal values within the interval, and the function
$Ra$ multiplied by the factor 20 to be seen on the scale.

The function $Ra$  is clearly observed to be a very slowly varying function of $r_\perp$ 
for all values of $x$ and $r_\perp$. Though at fixed $x$ 
the function $Ra$  cannot be  claimed to be exact constant, its variations with $r_\perp$ 
are very much suppressed comparing to the variations of the functions $Nr$ and $Ny$. 
For example, at $x=10^{-5}$ within the given interval the function $Ra$ changes by maximum
15\%. Within the very same interval both functions  $Nr$ and $Ny$ change in several
times. Then the relative fluctuation  is much less than 10\%, which
according to the condition (\ref{cond})  confirms the scaling. The phenomenon  
holds with a few percent   accuracy  and it
 improves at smaller $x\simeq10^{-7}$ and in the  deep saturation
region. However to observe  this scaling behavior 
in these regions  is numerically more problematic since 
both derivatives $Nr$ and $Ny$  tend to zero. 

As was discussed in the Introduction, we should expect the scaling violation at distances
of order $1/Q_s(x)$. The shorter distances are in the realm of applicability of the linear
equation which is not supposed to display any scaling phenomena. Nevertheless, we 
 observe   the scaling  
actually to exist also at distances which are much shorter than the saturation scale. The above 
statements seem to contradict  each other. We believe, however, that the resolution of the 
paradox is in the linear equation which in fact   
exhibits the  approximate scaling behavior \cite{LT1}. 
Unfortunately, this numerical coincident prevents us from determination of the saturation scale
as a scale of the scaling violation.

\subsection{Saturation scale}

\begin{figure}[htbp]
\begin{tabular}{c c c}
\epsfig{file=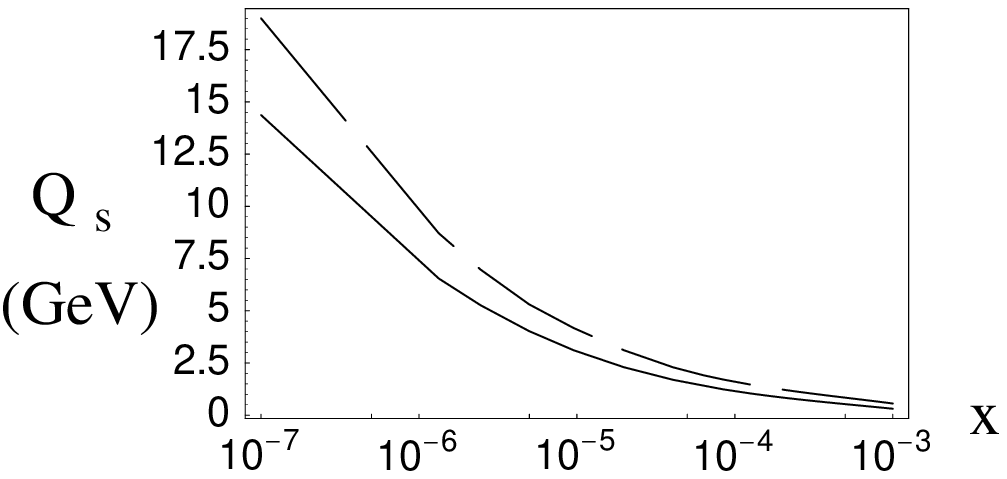,width=58mm, height=35mm} & 
\epsfig{file=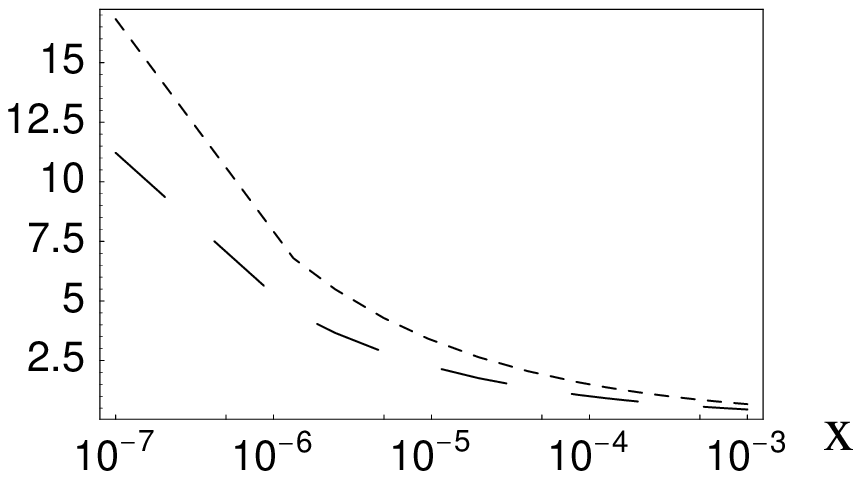,width=50mm, height=35mm} &
\epsfig{file=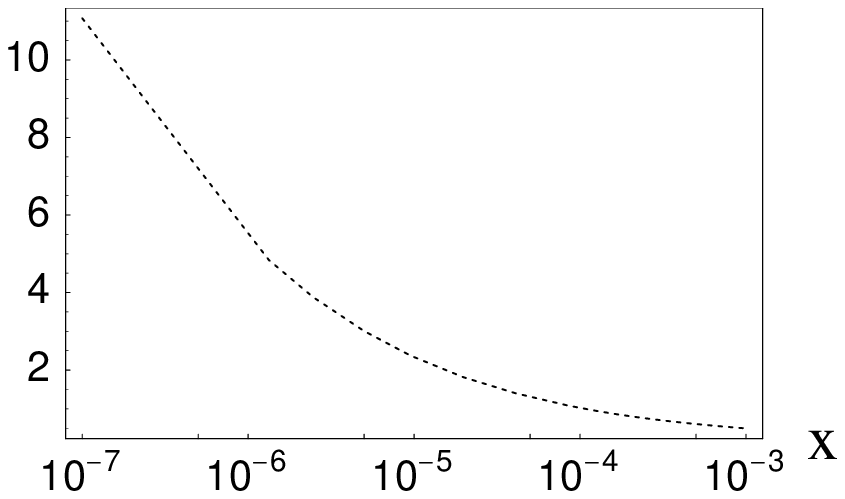,width=50mm, height=35mm}  \\
(a) & (b) & (c) \\
\end{tabular}
\caption[]{\it The saturation scale $Q_s$  is plotted as a function of $x$. 
(a) - the scales obtained from the equations (\ref{scale1}) (solid line)  and (\ref{scale2}) (dashed 
line), 
(b)  - the equation (\ref{Qs3}) is used to determine the scale.
(c) - the result obtained from  the equation (\ref{def4}).}
\label{scaleplot}
\end{figure}

No exact mathematical definition of the saturation scale is known so far. In the Ref. \cite{LGLM}
two definitions of the saturation scale were proposed and 
the solutions obtained from the equation (\ref{EQ}) 
were used  for its  determination.
 For the  step-like  function it is natural to define the saturation
scale  as a position where $\tilde N=1/2$:
\begin{itemize}
\item 
{\bf Definition 1:} 
\beq
\label{scale1}
\tilde N(2/Q_s, x)\,=\,1/2\,.
\eeq\end{itemize}
An alternative  definition of the saturation scale is:
\begin{itemize}\item
{\bf Definition 2:} 
\beq
\label{scale2}
\kappa \equiv\,-\,\ln[1\,-\,\tilde N(2/Q_s,x)]\,=\,1/2\,.
\eeq
\end{itemize}
The latter definition is related to the $b$-dependence of the solution and is 
motivated by the GM formula with $\kappa$ being the gluon packing 
factor\footnote{In the present paper as well as in the previous papers \cite{LGLM,LL} we 
do not deduce the saturation scale $Q_s$ but rather the dipole saturation radius $R_s$.  
The equality  $Q_s\equiv 2/R_s$ is motivated by the double logarithmic approximation. Though
for the equation (\ref{EQ}) this approximation is not justified, we still believe it to make reliable
estimates provided $Q_s$ is sufficiently large.}.
This definition is equivalent  to $\tilde N(2/Q_s,x) \simeq 0.4$ which  predicts
a somewhat larger saturation scale $Q_s(x)$ comparing with (\ref{scale1}).
The saturation scales obtained through the 
equations (\ref{scale1}) and (\ref{scale2}) are plotted in
the figure (\ref{scaleplot},a).

The saturation scale can be deduced directly from the scaling property (\ref{SCALING})
 which have been established.
To this goal  we can regard the equation (\ref{Ra}) as a new definition of the saturation scale:
\begin{itemize}
\item
{\bf Definition 3:} 
\beq
\frac{d \ln Q_s^2(x)}{d \ln x}\,=\,Ra(r_\perp,x)\,=\,const(r_\perp)
\label{def3}
\eeq
\end{itemize}
This definition allows us to determine the energy dependence of the saturation scale.
Note from the Fig. (\ref{ratio})  that 
the function $Ra$ is  practically independent of $x$, $Ra\simeq -0.7$. Hence we
obtain from  (\ref{def3}) the power law dependence of the saturation scale on $x$:
\beq
\label{Qs3}
Q_s(x)\,=\,Q_{s0}\,x^{-q}\,=\,Q_{s0}\,e^{-q \ln x}\,;\,\,\,\,\,\,\,
q\,=\,0.35\,\pm \,0.04\,.
\eeq
The obtained dependence of the saturation scale on $x$ is somewhat weaker than both the
double logarithmic prediction $q=2\as N_c/\pi$ \cite{LT1} and the numerical result of the 
Ref. \cite{Braun2} but significantly stronger  than
 the GW saturation model  $q_{GW}=0.288/2$ \cite{GW}.
Unfortunately, the parameter $Q_{s0}$ cannot 
be deduced from the scaling analysis only.  In order to make some estimates we choose
two reasonable values for $Q_{s0}$ just fixing the saturation scale at $x=10^{-4}$: 
$Q_s(10^{-4})=1\,GeV$ and $Q_s(10^{-4})=1.5\,GeV$. The obtained results are plotted
in the Fig. (\ref{scaleplot},b).

The physical meaning of the scale at which both $Nr$ and $Ny$ are extreme is
that at this scale (which is a function of $x$) the nonlinear effects responsible for the saturation
set in. Thus it is  natural to suggest yet another definition of the saturation scale
\begin{itemize}
\item 
{\bf Definition 4:}
\beq
\label{def4}
\left (\frac{\partial^2 \tilde N}{\partial r_\perp^2 \partial x}\right )_{r_\perp^2=4/Q_s^2(x)}\,
\simeq\, 
\frac{\partial}{\partial r_\perp^2} 
\left (r_\perp^2\,\frac{\partial \tilde N}{\partial r_\perp^2}\right)_{r_\perp^2=4/Q_s^2(x)}\,=\,0\,.
\eeq
\end{itemize}
It is important to stress that the definitions (\ref{def3}) and (\ref{def4}) are not equivalent,
and no one is a consequence of another.  However, if we suppose that 
$\kappa\sim (r_\perp^2)^{1-\gamma}$ in analogy with the GM
formula then  both definitions (\ref{def3}) and  (\ref{def4}) are equivalent.

The Fig. (\ref{scaleplot},c) shows the saturation scale $Q_s(x)$
obtained from (\ref{def4}) as a function of $x$. The values  presented are deduced 
with few percent errors.
The Fig. (\ref{scaleplot},c) predicts smaller
saturation scales compared with the ones from the Fig. (\ref{scaleplot},a). 
This fact can be naturally
explained if we suppose again that $\kappa\sim (r_\perp^2)^{1-\gamma}$.
Then it is easy to show that the definition (\ref{def4}) corresponds to the condition
$\kappa(x,2/Q_s)=1$, which implies $\tilde N(2/Q_s,x)=1-1/e \simeq 0.63$. Hence,
 the saturation is obtained at larger distances.

\section{Scaling phenomena for nuclei}

Solutions of the equation (\ref{EQ}) were obtained for  nucleus
 targets in the recent paper \cite{LL}.  
 All details about the solutions  for six nuclei $Au_{197}$, $Nd_{150}$,
$Mo_{100}$, $Zn_{70}$, $Ca_{40}$, and $Ne_{20}$ can be found there. 
In the present work  the only modification we perform is again 
concerns the large distance extrapolation of
the initial conditions. To this goal we use the Glauber formula for the initial conditions. For the 
nucleon cross section we use the result of the previous section.
This way we discover that the Glauber formula extrapolates the large distance behavior of the 
initial conditions consistently with both the asymptotics and  scaling.
The solutions obtained $\tilde N_A(r_\perp,x)$
 display  similar step-like behavior as in the proton case (Fig. \ref{sol}).

\begin{figure}[htbp]
\begin{tabular}{c c}
 \epsfig{file=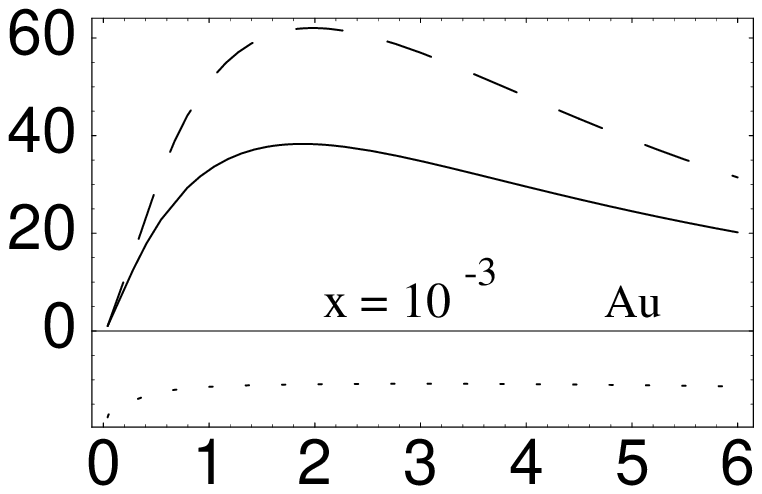,width=70mm, height=38mm}&
\epsfig{file=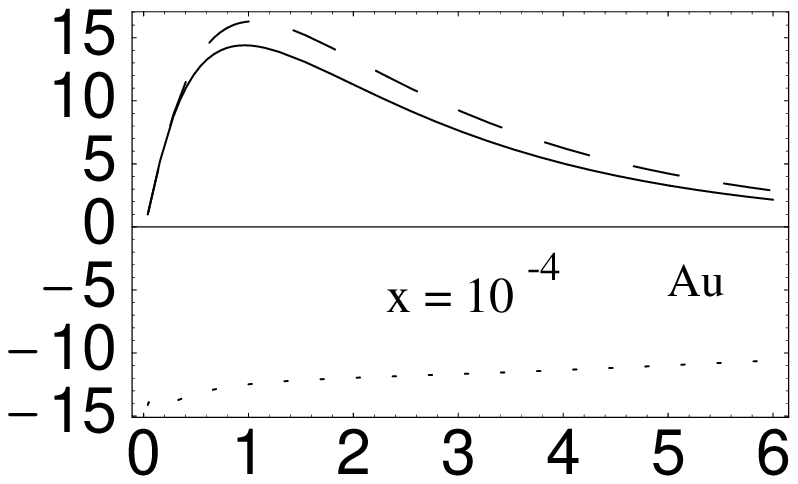,width=66mm, height=38mm}\\ 
 \epsfig{file=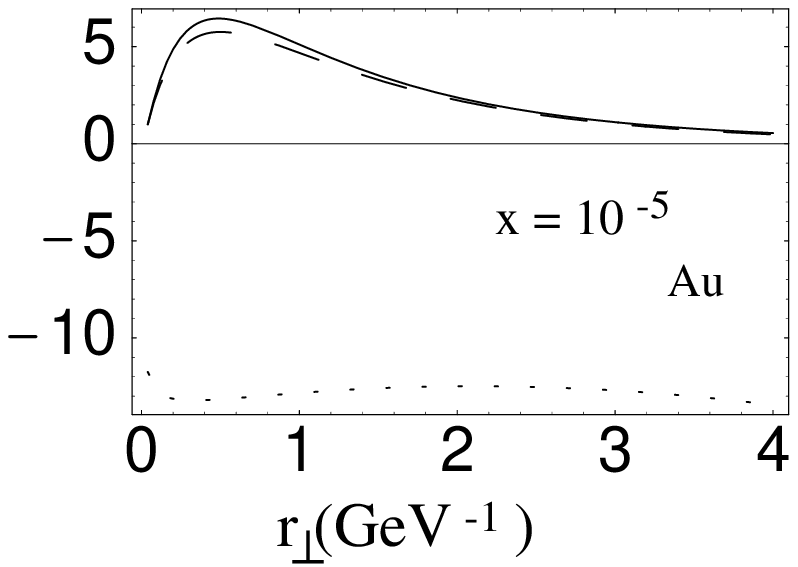,width=66mm, height=45mm}&
\epsfig{file=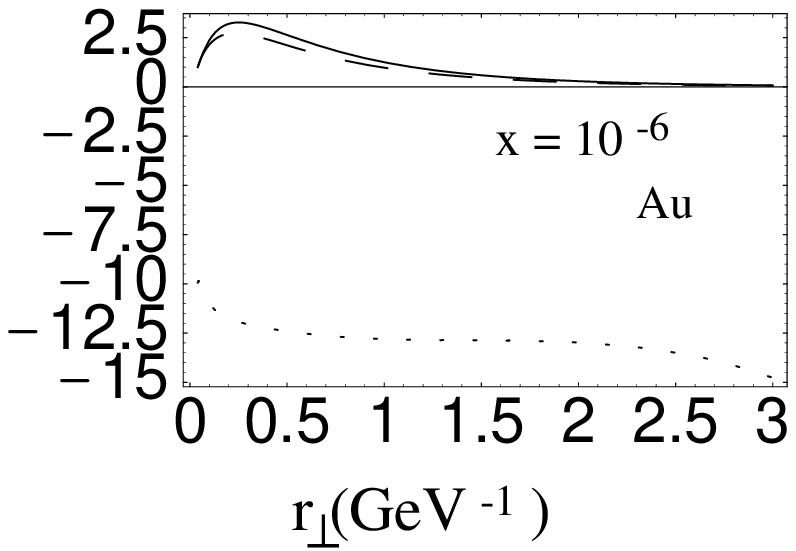,width=70mm, height=45mm}\\ 
\end{tabular}
  \caption[]{\it The  scaling   as a function of distance the $r_\perp$). 
The positive curves
  are $Nr_{Au}/Nr_{Au\,min}$ and   $Ny_{Au}/Ny_{Au\,min}$. The dotted line is 
$20\times Ra_{Au}$.}
\label{rationucl}
\end{figure}

In this section we investigate the scaling phenomena
for nuclei following the very same strategy as presented
above.
We start from computations of the derivative functions $Nr_A$ and $Ny_A$ for the nuclei, where
the subscript $A$ stands for  the atomic number of a nucleus. 
In fact, the dependences quite similar
to the Figs. (\ref{deriv}) and (\ref{ratio}) are obtained. The Fig. (\ref{rationucl}) presents an  
example of calculations  for the most heavy nucleus $Au$. Since in our approach the solution
for $Ne$ is almost identical to the proton (see the Ref. \cite{LL} for the discussion)  
   $Ne$ nucleus displays exactly the same scaling phenomena as proton. 

The scaling  on gold is observed form the Fig. (\ref{rationucl}). The other nuclei
display the very same phenomenon. Though for nuclei the numerical 
fluctuations are larger  it holds within at least 10\% accuracy with respect to the condition 
(\ref{cond}).  

The ratio $Ra_A(x)$  is almost $x$ independent with less than 20\% fluctuations. Moreover,
within the same accuracy  it is  $A$-independent function as well. Recalling the defenition
(\ref{def3}) of the saturation scale, we obtain
\beq
Q_s(x)\,=\,Q_{s0}(A)\,x^{-q_N}\,;\,\,\,\,\,\,\, \,\,\,\,\,\,\, q_N\,\simeq \,0.32\,\pm 0.05\,.
\eeq
The power $q_N$ is similar to the  power $q$ obtained for the proton.
Note that the $A$ dependence of the saturation scale is found to be $x$ independent:
$Q_{s0}(A)\sim A^{p_1}$, where $p_1$ is a constant. As was explained for the proton case,
the initial values $Q_{s0}(A)$ and hence the power $p_1$ cannot be deduced from the scaling
behavior only.

\begin{figure}[htbp]
\begin{tabular}{c c}
\epsfig{file=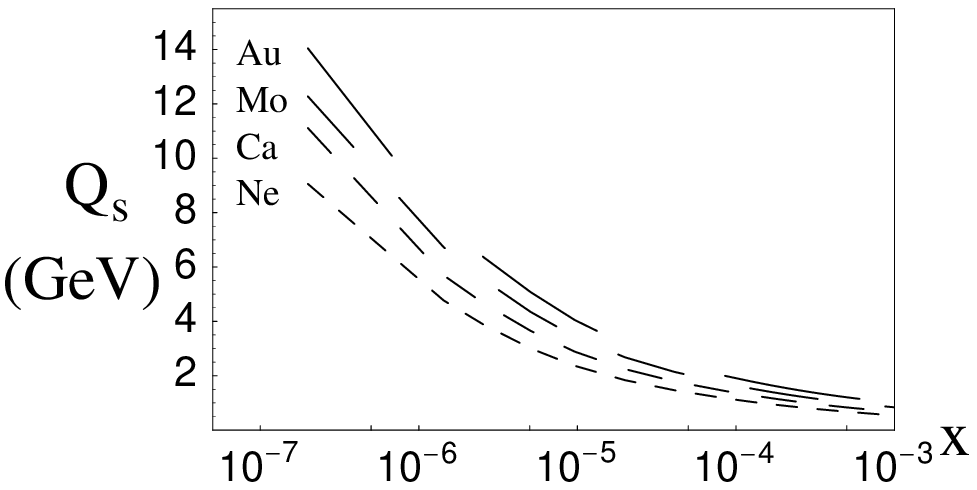,width=75mm, height=35mm} &  
\epsfig{file=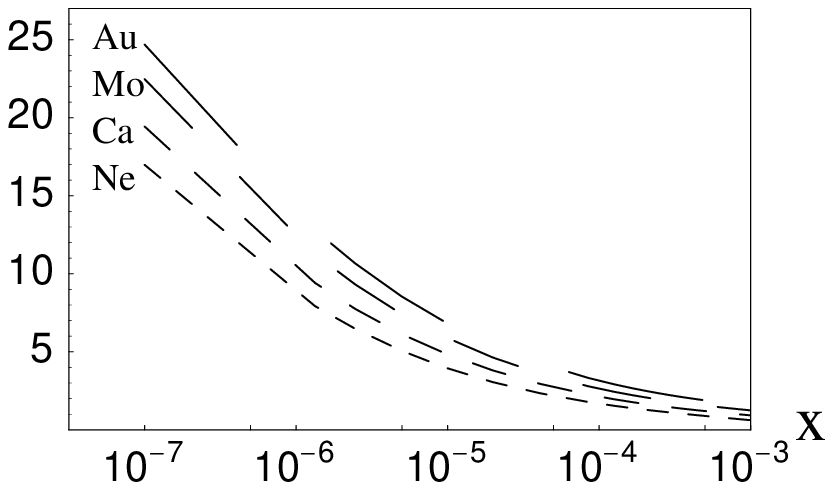,width=65mm, height=35mm} \\
(a)  & (b) \\
\end{tabular}
\caption[]{\it The saturation scale $Q_s$   
is plotted as a function of  $x$ for four nuclei $Au$, $Mo$, $Ca$, and $Ne$.
(a) - The result is obtained from (\ref{def4}); (b) - the result is obtained from  (\ref{scale2})
\cite{LL}.}
\label{scalnucl}
\end{figure}

The saturation scales according to the definition (\ref{def4}) are shown in the Fig.
(\ref{scalnucl},a) for four nuclei $Au$, $Mo$, $Ca$, and $Ne$. The numerical errors
do not exceed  10\%. For the sake of comparison we present in the Fig. (\ref{scalnucl},b)
the saturation scale obtained from (\ref{scale2}) \cite{LL}.
From the saturation scale obtained we can deduce its
 dependence on the atomic number $A$,
where the power law $Q_s\sim A^{p_2(x)}$ is assumed. All the nuclei are divided on two groups:
light nuclei ($Ne$, $Ca$, $ Zn$) and  heavy nuclei ($Zn$, $Mo$, $Nd$, and $Au$).
  The table (\ref{t1}) presents the powers
$p_2$ for various values of $x$.   The relative errors in the table are 
estimated not to exceed 20\%. On one hand,  the power $p_2$ is seen to decrease 
with decreasing $x$. This observation agrees with the results of the Ref. \cite{LL}. On the other
hand,  it can  be deduced from the table (\ref{t1}) that 
within the errors the power $p_2(x)$ can be viewed as a  constant
and its average value is in a perfect 
 agreement with the  value 2/9 -- the result of the Ref. \cite{Braun2}. 

\begin{table}
\begin{minipage}{9.0 cm}
\center{
\begin{tabular}{||l||c|c|c|c|c||} 
Nuclei $\,\,\backslash \,\,x$ & $10^{-7}$ & $10^{-6}$ & $10^{-5}$ & $10^{-4}$  & $10^{-3}$ \\
\hline \hline
 Light &                                       0.20    &       0.23  &    0.26   &     0.26 & 0.29        \\  
\hline
 Heavy &                                    0.18     &       0.18   &    0.22      &     0.22   & 0.26     \\
\hline
 All   &                                        0.19     &       0.20   &    0.23     &     0.23     & 0.27   \\
\hline
\end{tabular}}
\end{minipage}
\begin{minipage}{8. cm}
\caption{The power $p_2(x)$ for various  values of $x$.}
\label{t1}
\end{minipage}
\end{table}

The scaling phenomena described above reveals themselves in the energetic gain for
performing experiments on  heavy nuclei.
 The solution for one nuclei at given $x$ coincides with
the solution for another nuclei but at different $x$: 
\beq
\tilde N_{A_1}(r_\perp,x)\,\simeq\,\tilde N_{A_2}(r_\perp,\lambda(A_1,A_2)\, x)\,.
\label{lambda}
\eeq
The coefficient $\lambda$
occurs to be $x$ independent (for example $\lambda(Ne,Au)\simeq 5$).
The relation  (\ref{lambda}) is a consequence of the scaling phenomena implying
$Q_{s,A_1}(x)\simeq Q_{s,A_2}(\lambda(A_1,A_2)\, x)$. This  leads to the relation
\beq
\label{lambda1}
\left( \frac{A_1}{A_2}\right )^{p_2} \,\simeq\, \lambda^{-q_N}\,.
\eeq
For $Au$ and $Ne$ this relation gives $\lambda(Ne,Au) \simeq 5$ in total agreement
with the direct analysis of the solutions.

\section{Conclusions}

In the present paper the scaling phenomena in DIS were studied.  
The research concentrated on the nonlinear evolution equation (\ref{EQ})
governing the dynamics.  The solutions to this equation were
 recently found numerically in the Ref.
\cite{LGLM} for the proton target and in the Ref. \cite{LL} for the nuclei. 

A criteria for the scaling based on the solution of the nonlinear evolution was defined and
checked numerically. For the proton we  found scaling in all kinematic regions 
of study 
($10^{-7} \le x \le 10^{-2}$, $0.25\, {\rm GeV^{2}}\le Q^2 \le 2.5\times10^3\,  {\rm GeV^{2}}$)
and with a very good accuracy of order  few percents ($\le 5\%$).  The result is 
in agreement with the discovery of the scaling in the experimental data on the
structure function $F_2$ \cite{scalingexp}.
The scaling behavior is predicted to improve  at the LHC  and THERA energies. 
It is important to note that the scaling phenomena exist also at distances much shorter than 
the saturation scale \cite{LT1}.
At very short distances, where the linear evolution takes place, no scaling should
be observed. Nevertheless we found that this scaling do exist numerically with about 10\%
accuracy  and we are not able to detect its violation. 

\begin{figure}[htbp]
\begin{tabular}{c c}
\epsfig{file=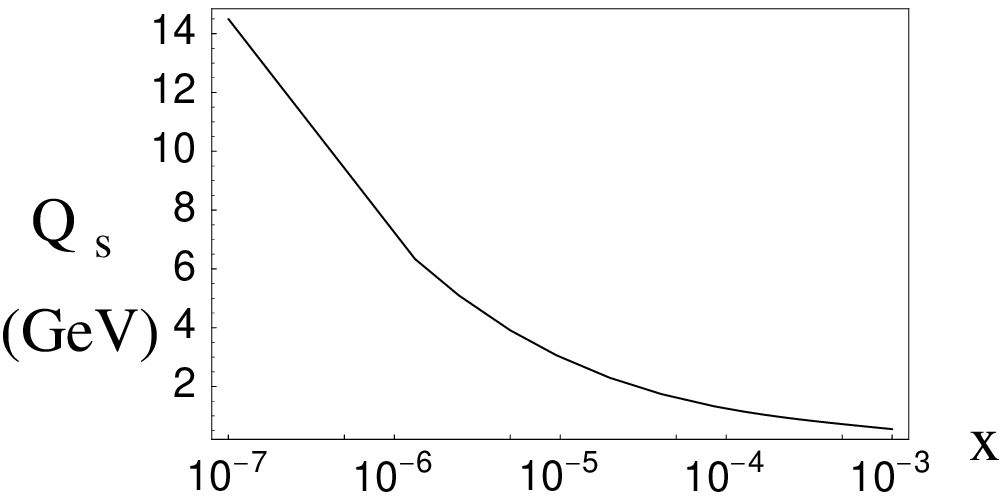,width=75mm, height=35mm} &
\epsfig{file=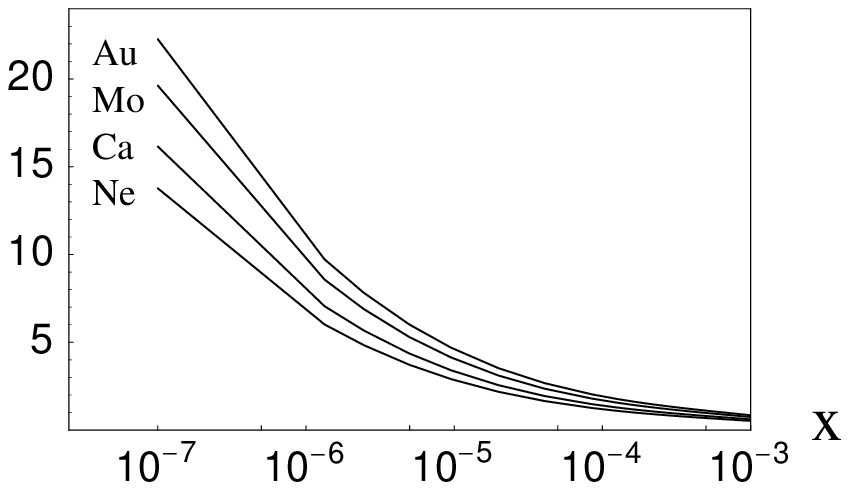,width=65mm, height=35mm} \\
(a) & (b) \\
\end{tabular}
\caption[]{\it The average saturation scale $Q_s$ as a function of $x$. (a) - the result for the 
proton;
(b) - the result for the nuclei. }
\label{scale_All}
\end{figure}

The  solution found in \cite{LGLM} of the nonlinear equation was used  to
estimate the saturation scale $Q_s(x)$.  In the present work we gave two new definitions
of the saturation scale based on the scaling phenomena. 
 In spite of considerable uncertainty in the value of the
 saturation scale  all the definitions  predict that it grows with decreasing $x$  
in accordance with the theoretical  expectations \cite{MV,GLR,MUQI,SAT}. If we 
allow  ourselves to average between all the result for the saturation scale depicted on the
figure (\ref{scaleplot})  we would obtain  a prediction shown in the  Fig. (\ref{scale_All},a). 
The relative errors for the latter are roughly 30\% for all $x$ which indicate the uncertainty
in the saturation scale definition.

The scaling phenomena were observed for the nuclear targets, which confirm conclusions
of the Ref. \cite{Braun2}. The saturation scale estimated  from the scaling  displays the power
law dependence on both the atomic number $A$ and the energy variable $x$:
 $Q_{s,A}(x)\sim A^{p_2}\,x^{-q_N}$. The value obtained for the power $p_2$ is in a good
agreement with the value 2/9  deduced in the Ref. \cite{Braun2}. 

Both the values of the nucleus saturation scales and their $A$ dependence obtained in the 
present paper are slightly different from the ones found in the Ref. \cite{LL}. The main source
of this effect  is certainly comes from the difference in the saturation scale definitions. Since
we do not know what definition is better we combine all the information we have, 
proceeding  similarly to the proton case. The results of this procedure are presented in the Fig.
(\ref{scale_All},b).  We hope  that our  predictions of the saturation scales  $Q_{s,A}$ 
for various nuclei  will serve   as a theoretical ground  for the 
 RHIC data analysis in high parton density QCD  \cite{RHIC}.

{ \bf Acknowledgments:} 

I wish to thank Eugene Levin for his inspiration and support of this work.  
 I am also  very grateful to E. Gotsman,   U. Maor,   
  and K. Tuchin  for illuminating discussions  on the subject.

\end{document}